\documentclass[12pt]{article}

\usepackage{amsmath,amssymb,amsfonts,amsbsy}
\usepackage{cite}
\usepackage{graphicx}
\usepackage{wrapfig}
\usepackage{epsfig}

\usepackage{bbm} 
\usepackage{bm} 
\usepackage{color}                                                       %
\usepackage{dsfont} 
\usepackage{latexsym} 
\usepackage{lscape} 
\usepackage{mathrsfs} 
\usepackage{morefloats} 
\usepackage{slashed} 
\usepackage{psfrag}

\def\xb{\overline{x}}

\def\als{\alpha_s}
\def\vk{{\bf k}_{\perp}}
\def\vbs{{\bf b}}

\begin{document}
\addcontentsline{toc}{subsection}{{Role of transversity in spin effects in
meson leptoproduction.}\\
{\it S.V. Goloskokov}}

\setcounter{section}{0}
\setcounter{subsection}{0}
\setcounter{equation}{0}
\setcounter{figure}{0}
\setcounter{footnote}{0}
\setcounter{table}{0}

\begin{center}
\textbf{ROLE OF TRANSVERSITY IN SPIN EFFECTS IN
MESON LEPTOPRODUCTION.}

\vspace{5mm}

S.V. Goloskokov$^{ \dag}$

\vspace{5mm}

\begin{small}
   \emph{Bogoliubov Laboratory of Theoretical Physics, Joint
Institute for
Nuclear Research, Dubna 141980, Moscow region, Russia} \\
  $\dag$ \emph{E-mail: goloskkv@theor.jinr.ru}
\end{small}
\end{center}

\vspace{0.0mm} 

\begin{abstract}
We analyze the light meson leptoproduction within the handbag
approach. We show that effects determined by the transversity
Generalized Parton Distributions (GPDs), $H_T$ and $\bar E_T$ are
essential in the description of pseudoscalar and vector meson
leptoproduction.
\end{abstract}

\vspace{7.2mm}

\section{Introduction}
\label{intro}

In our  papers \cite{gk06}, we  calculated the processes of light
meson leptoproduction within the handbag approach, where the
amplitudes factorize into hard subprocesses and in  (GPDs)
\cite{fact} which encode  soft physics.  The modified perturbative
approach \cite{sterman}, where the quark transverse degrees of
freedom accompanied by Sudakov suppressions are taken into
account, was used to calculate  the hard subprocess amplitudes. We
discuss some details of this approach for vector meson (VM)
production in  section 2.

The pseudoscalar meson (PM) production was analyzed in \cite{gk09,
gk11}. It was found that the transversity GPDs $H_T$ and $\bar
E_T$ are essential in the description of these reactions at low
$Q^2$. Within the handbag approach the transversity GPDs are
accompanied by  twist-3 meson distribution amplitudes. These
transversity contributions provide large transverse cross sections
for most of the pseudoscalar meson channels \cite{gk11} (see
section 3)

The role of transversity GPDs in the VM leptoproduction
\cite{gk13} is discussed in section 4. The importance of the
transversity GPDs was examined in  the Spin Density Matrix
Elements (SDMEs) and in asymmetries measured with a transversely
polarized target. For the transversity GPDs $H_T$ and $\bar{E}_T$
we used the same parameterizations as in our study of the PM
leptoproduction. Our results for SDMEs are in good agreement with
HERMES experimental data on the $\rho^0$ production. We also
estimated the moments of transverse target spin asymmetries
$A_{UT}$ which contain the transversity contributions. The
 $A_{UT}^{\sin (\phi_s)}$ asymmetry is found to be not small \cite{gk13} at
COMPASS energies.

\section{Meson leptoproduction and handbag approach}
The  amplitude of meson leptoproduction at large $Q^2$ is assumed
to factorize  \cite{fact} into a hard subprocess amplitude ${\cal
H}$ and a soft proton matrix element, parameterized in terms of
GPDs $F(\xb,\xi,t),  E(\xb,\xi,t), ...$.

The proton non-flip and spin-flip amplitude can be expressed in
terms of gluons, quarks or sea contributions
\begin{equation}\label{ff}
 {\cal M}_{\mu' +,\mu +} \propto \int_{-1}^1 dx
   {\cal H}^a_{\mu' +,\mu +} F^a(x,\xi,t),\;
{\cal M}_{\mu' -,\mu +} \propto \frac{\sqrt{-t}}{2 m}\int_{-1}^1
dx
   {\cal H'}^a_{\mu' +,\mu +} E^a(x,\xi,t).
\end{equation}

The subprocess amplitude is calculated within the MPA
\cite{sterman}. The  amplitude ${\cal H}^a$ is   a contraction of
the hard part ${\cal F}^a$ which includes the transverse quark
momentum $\vk$ in the propagators and the nonperturbative  meson
wave function $\Psi(\vk)$ \cite{koerner}. The gluonic corrections
are treated in the form of the Sudakov factors. The resummation
and exponentiation of the Sudakov corrections $S$ can be performed
in the impact parameter space $\vbs$ \cite{sterman}, and the
amplitude reads as
\begin{eqnarray}\label{aa}
{\cal H}^a_{0\lambda,0\lambda}\propto \int d\tau d^2b\,
         {\Psi}(\tau,-\vbs)\,\nonumber
         {\cal F}^{a}_{0\lambda,0\lambda}(\xb,\xi,\tau,Q^2,\vbs,)\,
      \als\, {\rm
      exp}{[-S(\tau,\vbs,Q^2)]}.
\end{eqnarray}
Here $\tau$ is the momentum fraction of the quark  that enters
into the meson.

The GPDs contain extensive information about the hadron structure.
Hadron form factors and parton angular momenta can be related with
GPDs. At zero skewness $\xi$ and momentum transfer GPDs are equal
to ordinary PDFs
\begin{equation}\label{pdf}
 F^a(x,0,0)=f^a(x),\;\; E^a(x,0,0)=e^a(x)
\end{equation}
Here quarks(valence and sea) and gluon PDFs $f^a$ are determined
from CTEQ6 parameterization \cite{CTEQ6}. The PDFs $e^a$ are taken
from the Pauli form factor \cite{pauli}.

 The GPDs are estimated using the double distribution
representation \cite{mus99} which connects  GPDs with PDFs through
the double distribution function $\omega$. For the valence quark
contribution it looks like
\begin{eqnarray}\label{ddf}
\omega_i(x,y,t)= h_i(x,t)\,
                   \frac{3}{4}\,
                   \frac{[(1-|x|)^2-y^2]}
                           {(1-|x|)^{3}}.
\end{eqnarray}
The functions $h$  are determined in the terms of PDFs and
parameterized in the form
\begin{equation}\label{pdfpar}
h(x,t)= N\,e^{b_0 t} x^{-\alpha(t)}\,(1-x)^{n}.
\end{equation}
Here  the $t$- dependence is considered in a Regge form and
$\alpha(t)$ is the corresponding Regge trajectory. The parameters
in (\ref{pdfpar}) are obtained from the known information about
PDFs e.g, \cite{CTEQ6, pauli}.
\begin{figure}[h!]
\begin{center}
\begin{tabular}{cc}
\includegraphics[width=6.1cm,height=5cm]{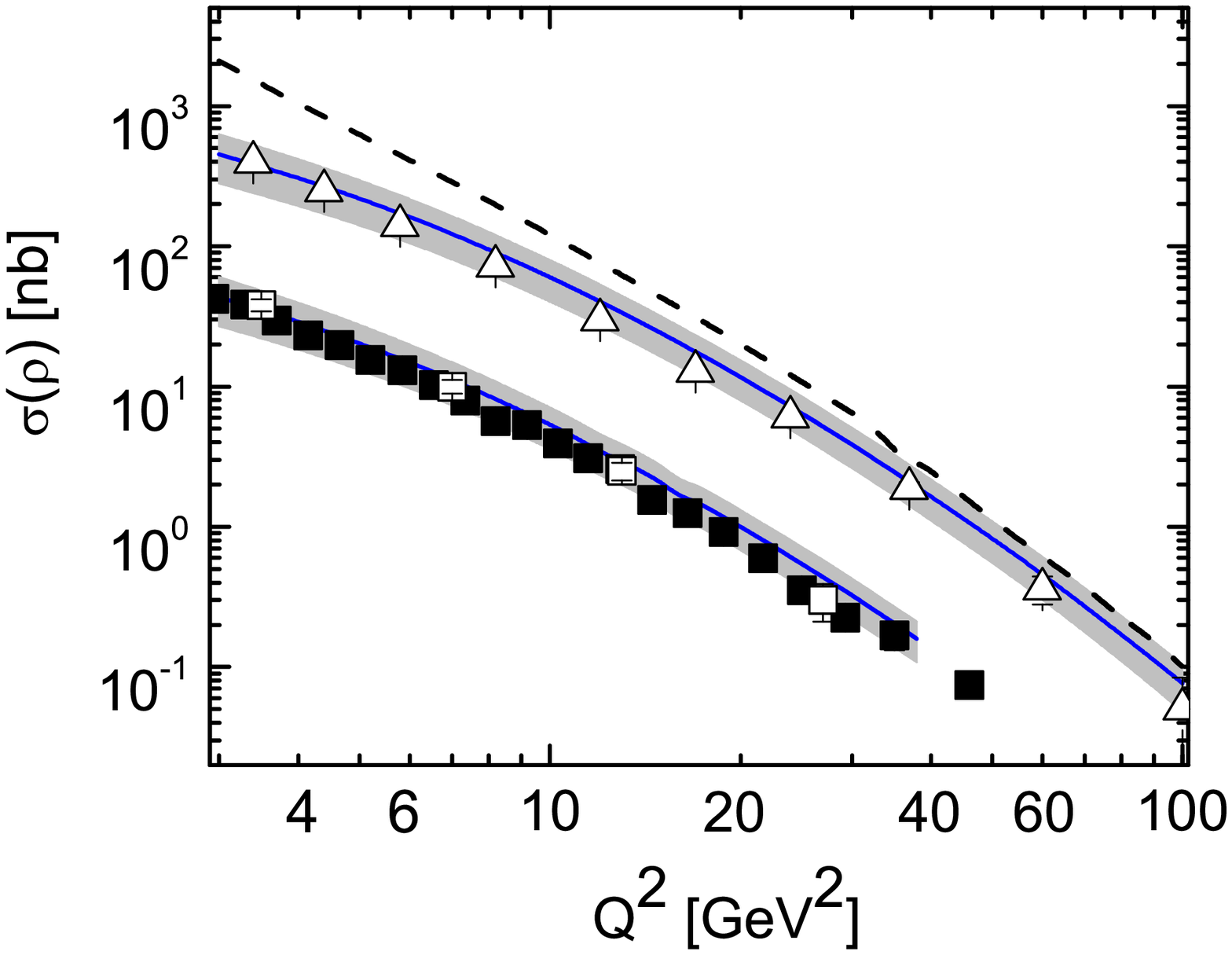}&
\includegraphics[width=6.1cm,height=5cm]{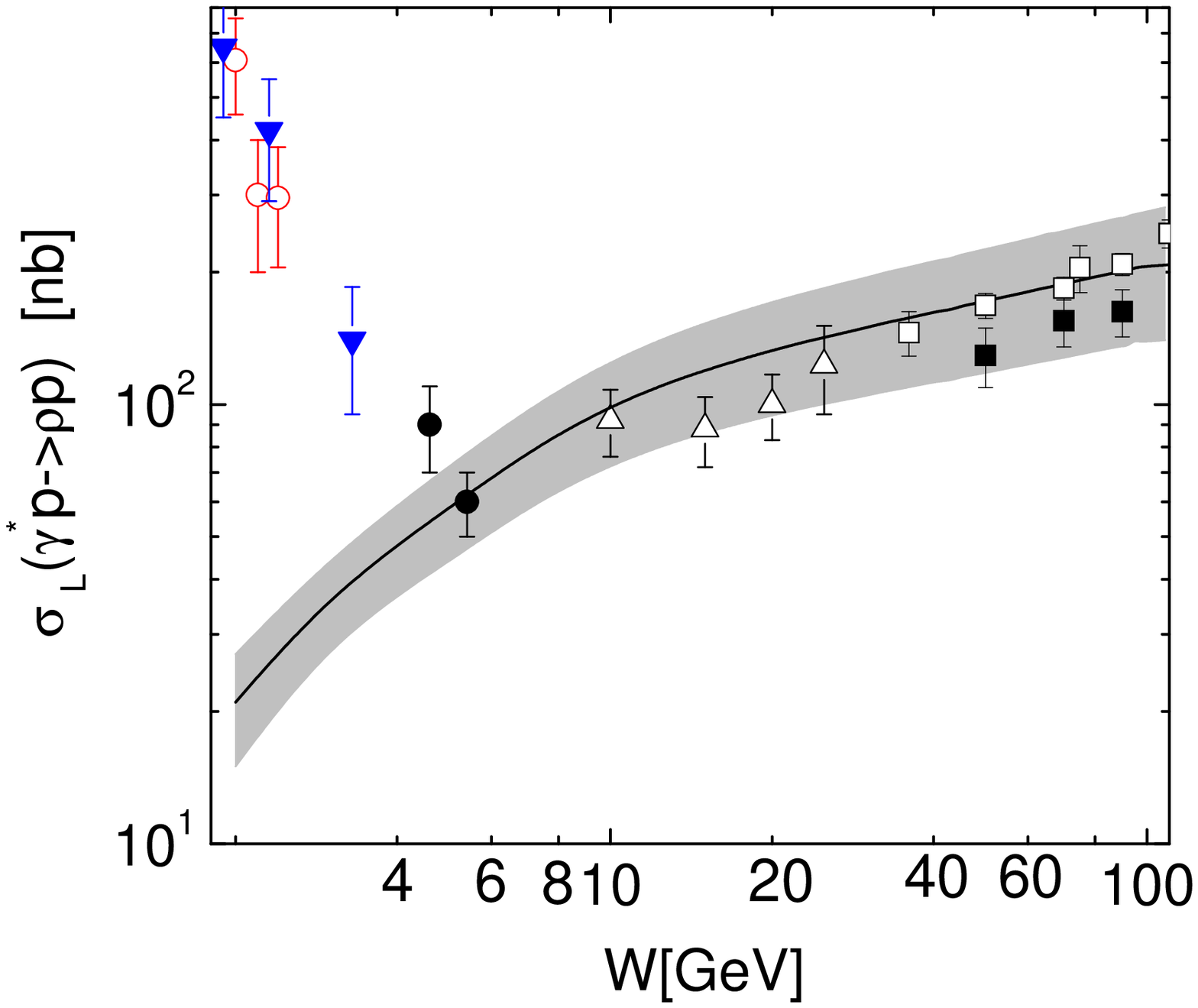}
\end{tabular}
\end{center}
\caption{Left: Cross sections of the $\rho$ production  at $W=75
\mbox{GeV}$/10 and $W=90 \mbox{GeV}$. Dashed line: leading twist
results. Right: The longitudinal cross section for the $\rho^0$
production at $Q^2=4.0\,\mbox{GeV}^2$. References to experimental
data can be found in \cite{gk06}}
\end{figure}

The handbag approach was successfully applied to light meson
leptoproduction \cite{gk06}. In Fig.1, we show our results for
$Q^2$ and $W$ dependencies of the $\rho$ leptoproduction which are
in  good agreement with experimental data. It can be seen in Fig.
1, (left) that the leading twist results do not reproduce data at
low $Q^2$. The power $k_\perp^2/Q^2$ corrections in the
propagators of hard   subprocess amplitude are important in the
description of the data. Corrections can be regarded as effective
consideration of the higher twist effects. From Fig 1 (right) we
see that the model describes the $\rho$ meson leptoproduction
quite well for $W> 4\mbox{GeV}$. The rapid growth of the cross
section at lower energies has not been understood within the
handbag model till now.

\section{Transversity in pseudoscalar mesons production}
Exclusive electroproduction of PM was studied  within the handbag
approach \cite{gk09, gk11}. It was shown that the asymptotically
dominant leading-twist contributions, which are determined by the
GPDs $\widetilde H$ and $\widetilde E$, are not suffcient to
describe the experimental results on electroproduction of PM at
low $Q^2$.  It can be seen, for example, from
$A_{UT}^{\sin(\phi_s)}$ asymmetry
\begin{equation}\label{aut}
A_{UT}^{\sin(\phi_s)} \propto \mbox{Im}[ M^*_{0-,++} M_{0+,0+}].
\end{equation}
This asymmetry was found to be small in the handbag model based on
the leading twist amplitudes. This result is inconsistent with the
data where $A_{UT}^{\sin(\phi_s)} \sim 0.5$.

 A new twist-3 contribution to the $M_{0-,++}$
amplitude, which is not small at $t' \sim 0$, is needed to
understand the data. The inclusion  in our consideration of the
$M_{0+,++}$ amplitude which has a similar twist-3 nature is also
extremely important  to explain the PM production at  low $Q^2$.
We estimate these contributions  by the transversity GPD $H_T$,
$\bar E_T$ in conjugation with the twist-3 pion wave function in
the hard subprocess amplitude ${\cal H}_{ 0-,\mu+}$ \cite{gk11}
\begin{equation}\label{ht}
{\cal M}^{M,tw-3}_{0-,\mu+} \propto \,
                            \int_{-1}^1 d\xb
   {\cal H}_{0-,\mu+}(\xb,...)\,H^{M}_T;\;
   {\cal M}^{M,tw-3}_{0+,\mu+} \propto \, \frac{\sqrt{-t'}}{4 m}\,
                            \int_{-1}^1 d\xb
 {\cal H}_{0-,\mu+}(\xb,...)\; \bar E^{M}_T.
\end{equation}

The $H_T$ GPD is connected with transversity PDFs  as
\begin{equation}
  H^a_T(x,0,0)= \delta^a(x);\;\;\; \mbox{and}\;\;\;
\delta^a(x)=C\,N^a_T\, x^{1/2}\, (1-x)\,[q_a(x)+\Delta q_a(x)].
\end{equation}
We parameterize the PDF $\delta$ (see \cite{gk09, gk11}) by using
the model \cite{ans}. The double distribution (\ref{ddf}) is used
to calculate GPD $H_T$.

At the moment, the information on $\bar E_T$ is very poor. Some
results were obtained only in the lattice QCD \cite{lat}. The
lower moments of $\bar E_T^u$ and $\bar E_T^d$ were found to be of
the same sign, similar in size and quite large. At the same time,
$H_T^u$  and $H_T^d$ have different signs.
\begin{figure}[h!]
\begin{center}
\begin{tabular}{cc}
\includegraphics[width=6.4cm,height=5cm]{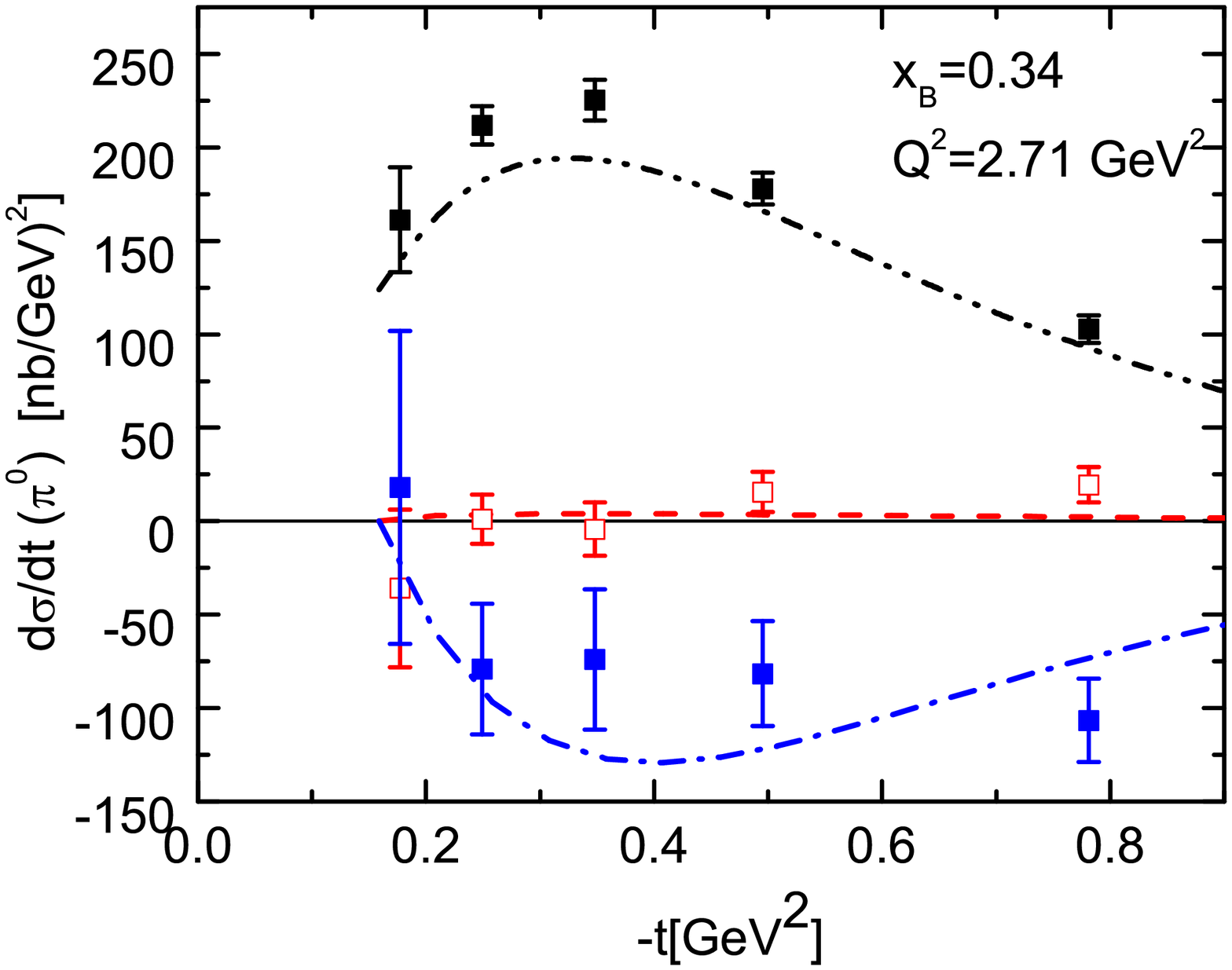}&
\includegraphics[width=6.1cm,height=5.1cm]{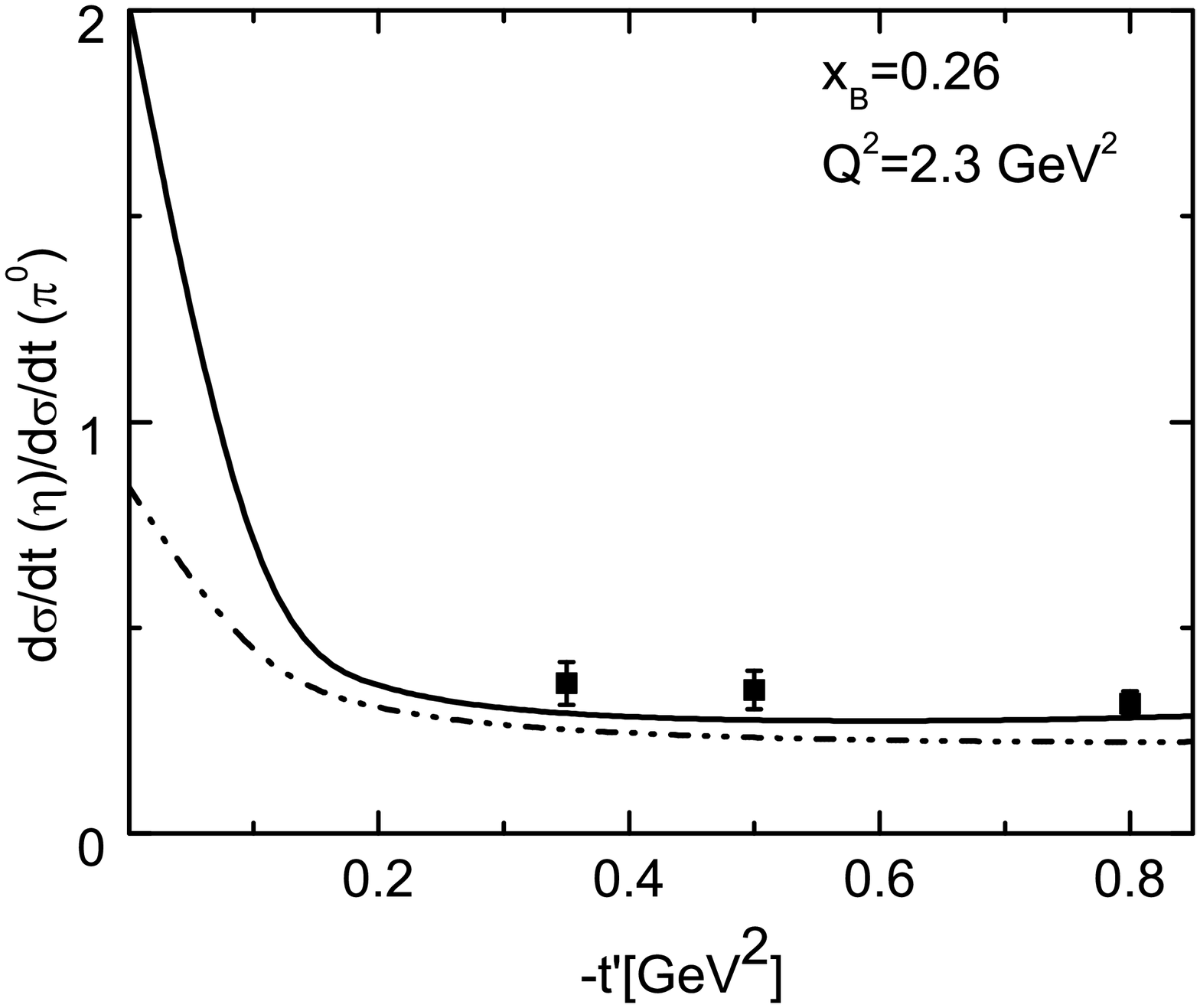}
\end{tabular}
\end{center}
\caption{Left: $\pi^0$ production in the CLAS energy range
together with the data \cite{bedl}. Dashed-dot-dotted line-
$\sigma=\sigma_{T}+\epsilon \sigma_{L}$, dashed
line-$\sigma_{LT}$, dashed-dotted- $\sigma_{TT}$. Right:
$\eta/\pi^0$ production ratio in the CLAS energy range together
with preliminary data \cite{vkubar}.}
\end{figure}
These properties of GPDs provide  essential compensation of the
$\bar E_T$ contribution in the $\pi^+$ amplitude, but $H_T$
effects are not small there. For the $\pi^0$ production we have
the opposite case -- the $\bar E_T$ contributions are large and
the $H_T$ effects are small.

 In Fig. 2 (left), we present our results for the cross section
of the $\pi^0$ production. The transverse cross section where the
$\bar E_T$ and $H_T$ contributions are important \cite{gk09}
dominates. At small momentum transfer the $H_T$ contribution is
visible and provides a nonzero cross section. At $-t' \sim 0.2
\mbox{GeV}^2$ the $\bar E_T$ contribution becomes essential and
gives a maximum in the cross section. A similar contribution from
 $\bar E_T$ is observed in the interference cross section
$\sigma_{TT}$. The fact that we describe well both unseparated
$\sigma$ and $\sigma_{TT}$ cross sections can indicate that
transversity effects were probably observed in CLAS \cite{bedl}.
In Fig. 2 (right), we show  the $\eta$ and $\pi^0$ cross section
ratio obtained in the model (for details see \cite{gk11}). At
small momentum transfer this ratio is controlled by the $H_T$
contribution. At larger $-t$ the $E_T$ contributions become
important. The value  about 1/3 for the cross section ratio in the
momentum transfer $-t'> 0.2\mbox{GeV}^2$ is a consequence of the
flavor structure of the $\eta$ and $\pi^0$ amplitudes. This result
was confirmed by the preliminary CLAS data \cite{vkubar}.
\begin{figure}[h!]
\begin{center}
\begin{tabular}{cc}
\includegraphics[width=6.1cm,height=5cm]{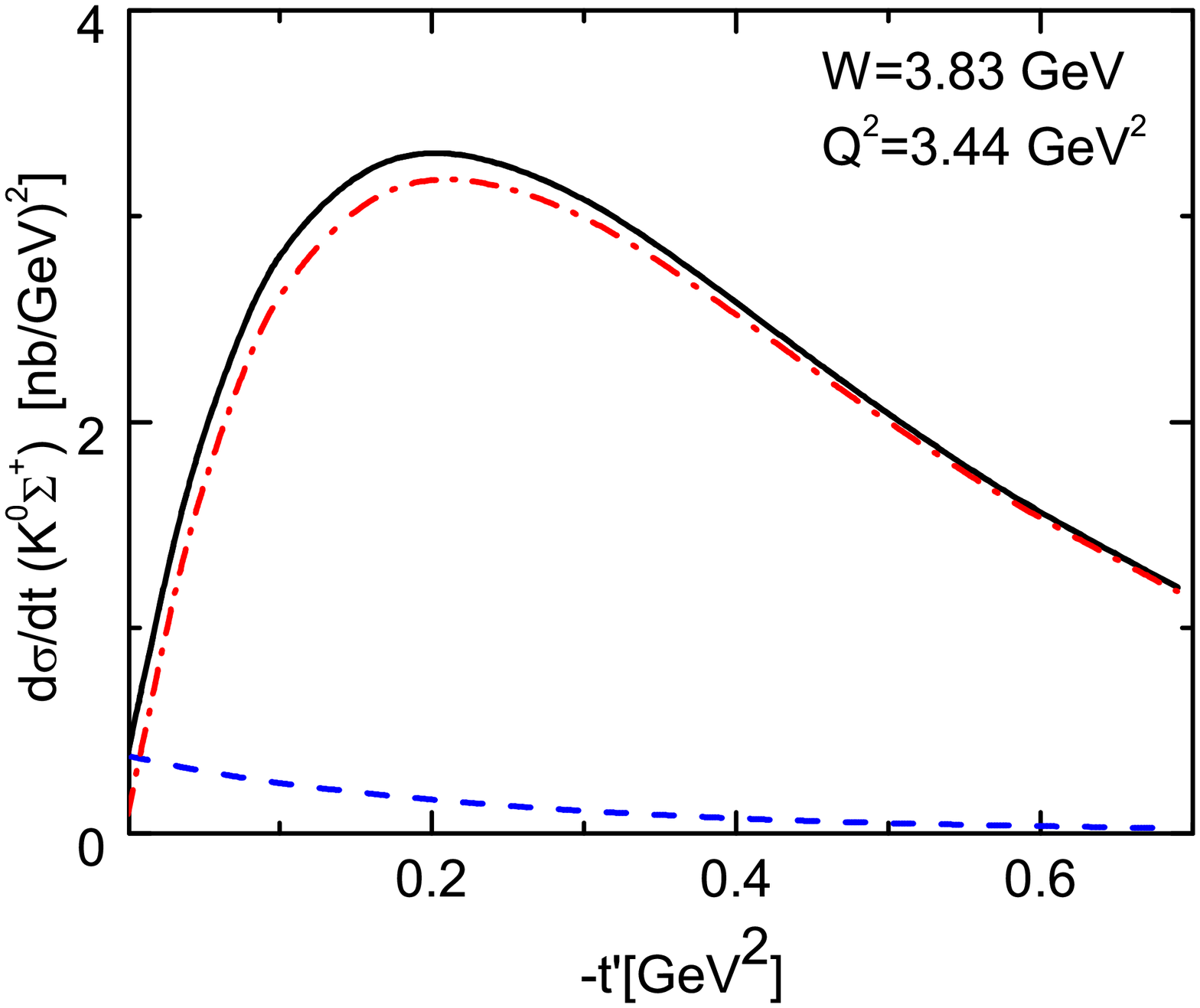}&
\includegraphics[width=6.1cm,height=5cm]{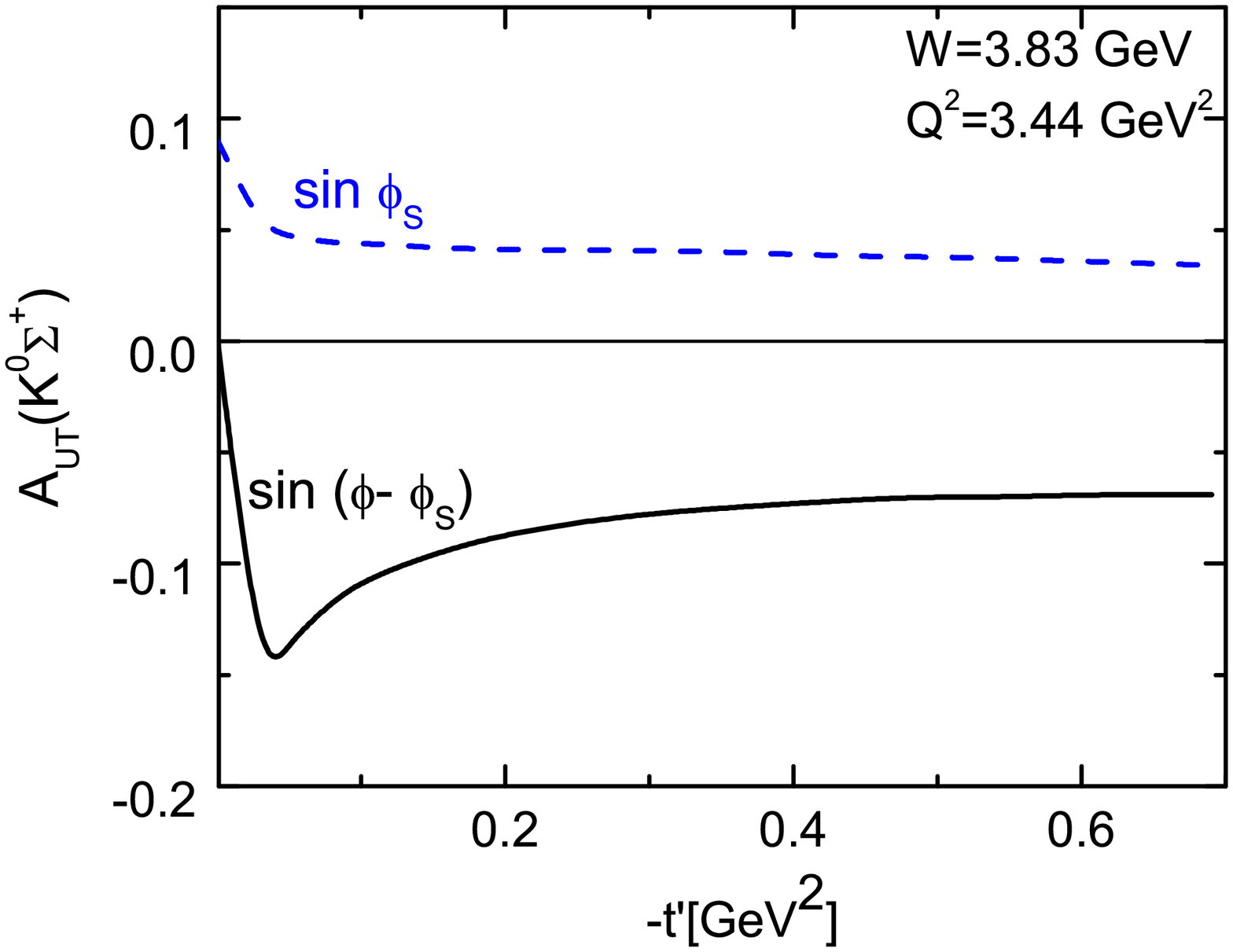}
\end{tabular}
\label{fig:5}
\end{center}
\caption{Left:  Cross sections of the $K^0 \Sigma^+$ production at
HERMES energies. Right: Predicted  moments of $A_{UT}$ asymmetries
for the $K^0 \Sigma^+$ channel at HERMES.}
\end{figure}

A similar essential transversity $E_T$ contribution is observed in
the kaon production. An example of our results for the $K^0
\Sigma^+$ cross section is shown in Fig. 3 (left). As in the
$\pi^0$ production, we find here a dip near $-t'=0$. It was found
that the longitudinal cross section $\sigma_L$, which is expected
to play an important role, is much smaller with respect to the
transverse cross section $\sigma_T$ at low $Q^2$- see Fig 3
(left).  At sufficiently large $Q^2$ the leading-twist $\sigma_L$
contribution will dominate because transversity twist-3 effects,
which contribute to $\sigma_T$, decrease quickly with $Q^2$
growing. The same result was found in the $\pi^0$ production
\cite{sgspin11}. The predicted asymmetries in $K^0 \Sigma^+$
channel are shown in Fig. 3 (right).

\section{Transversity in vector mesons production}
Now we  extend our analysis of transversity effects to theVM
production \cite{gk13}. Transversity will be essential in the
amplitudes with a transversely polarized photon and a
longitudinally polarized vector meson. The twist-3 amplitudes have
a form of (\ref{ht}) where the transversity GPDs occur in
combination with twist-3 meson wave functions. The asymptotic form
for the twist-3 chiral-odd DA $h^{(s)}_{||V}=6 \tau (1-\tau)$ is
used.

Note that the transversity contribution in the VM production
contains the parameter $m_V=0.77 \mbox{GeV}$ instead of $\mu_\pi=2
\mbox{GeV}$ for PM production \cite{gk13}. As a result, the
transversity contribution to the VM amplitudes is parametrically
about 3 times smaller with respect to PM case. In calculation of
the amplitude we use the same parameterizations for transversity
GPDs $H_T$ and $\bar E_T$ which was obtained in our study of the
PM leptoproduction in the section 3.
\begin{figure}[h!]
\begin{center}
\begin{tabular}{cc}
\includegraphics[width=6.1cm,height=5cm]{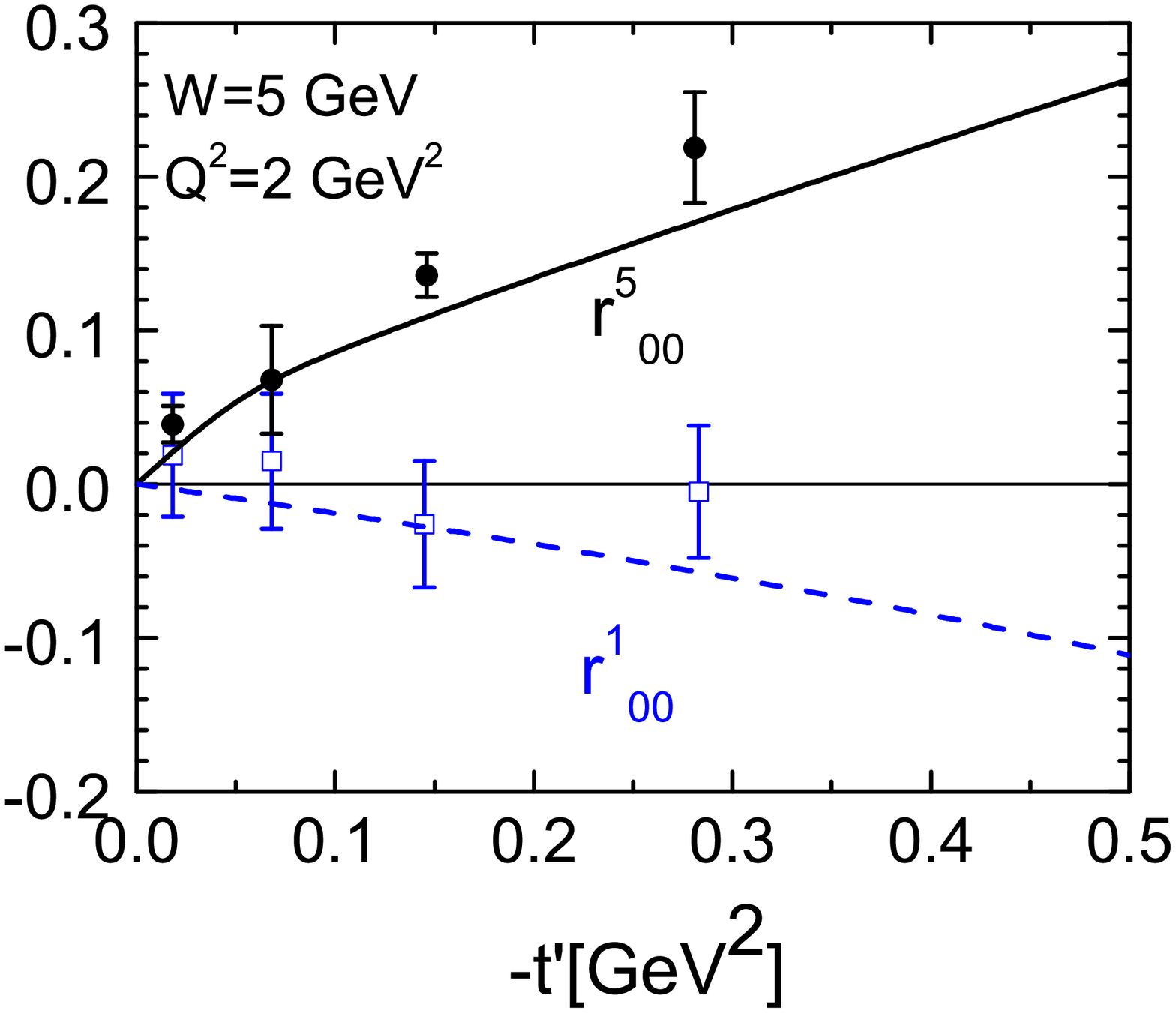}&
\includegraphics[width=6.1cm,height=5cm]{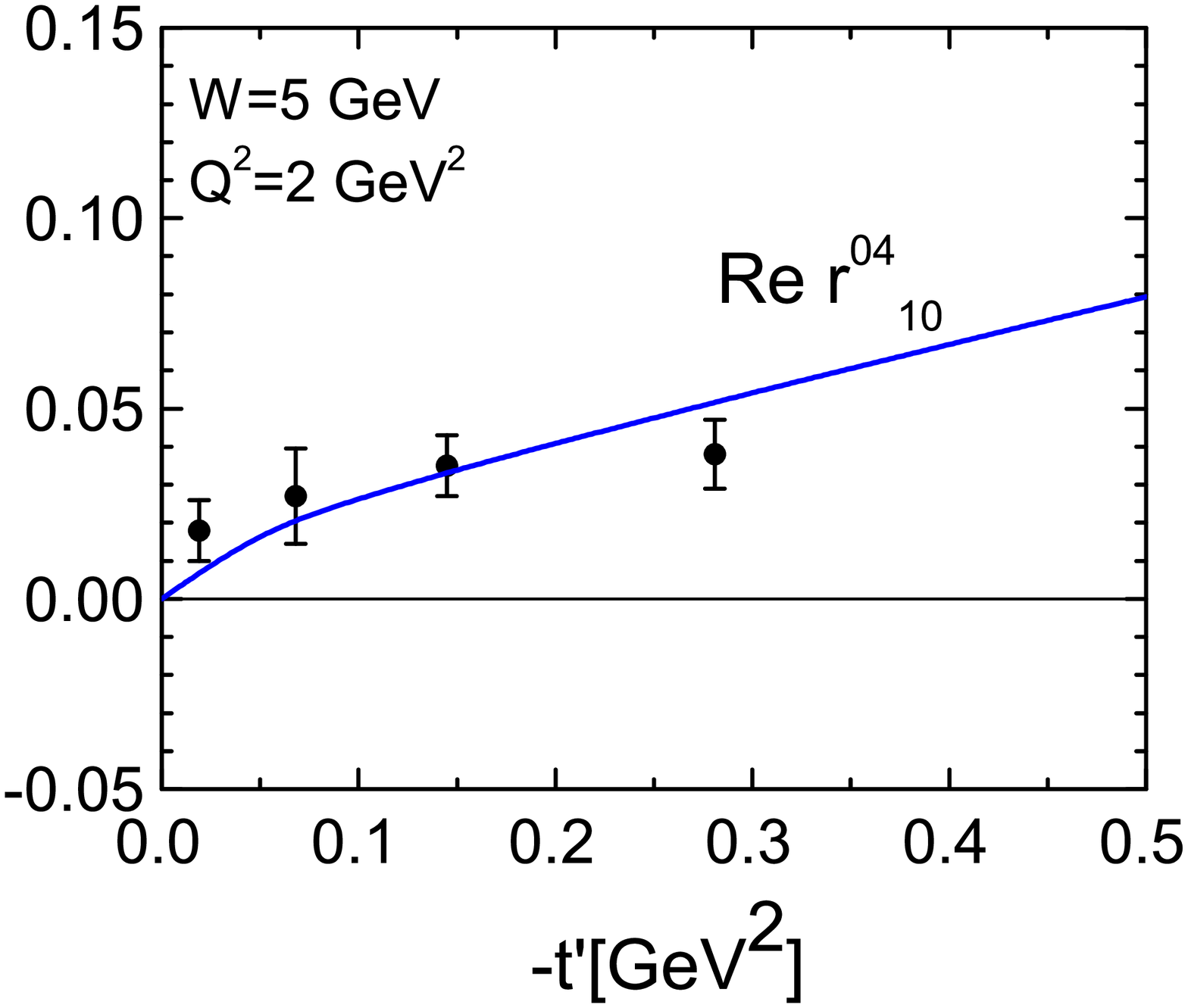}
\end{tabular}
\end{center}
\caption{Transversity effects at SDMEs at $W=5 \mbox{GeV}$
together with HERMES data \cite{airap}.}
\end{figure}

 The importance of the transversity GPDs was examined
in  the  SDMEs and in asymmetries measured with a transversely
polarized target. The $M_{0+,++}=<\bar E_T>$ amplitude is
essential in some SDMEs. Really,
\begin{equation}\label{sdme}
r^5_{00} \sim \mbox{Re}[M_{0+,0+}^* M_{0+,++}];\;\;\; r^1_{00}
\sim -|M_{0+,++}|^2;\;\;\;r^{04}_{10} \sim \mbox{Re}[M_{++,++}^*
M_{0+,++}].
\end{equation}

Our results   for these the SDMEs  in the $\rho^0$ meson
production at HERMES are shown in  Fig. 4. These values and signs
are in good agreement with HERMES experimental data \cite{airap}.
We observe that large $\bar E_T$ effects found in the $\pi^0$
channel are compatible with SDME of the $\rho$ production at
HERMES energies.

In Fig. 5, we show  our results for the $\sin(\phi-\phi_s)$ moment
of the $A_{UT}$ asymmetry
\begin{equation}\label{autfmfs}
  A_{UT}^{\sin(\phi-\phi_s)} \sim \mbox{Im}[M_{0-,0+}^* M_{0+,0+} -
M_{0-,++}^* M_{0+,++}]
\end{equation}
at HERMES and COMPASS energies. This asymmetry is determined
essentially by interference of the $<\bar E>$ and $<F>$
contributions (\ref{ff}) and is consistent with the data. The
effects of transversity  are quite small here.

\begin{figure}[h!]
\begin{center}
\begin{tabular}{cc}
\includegraphics[width=6.1cm,height=5cm]{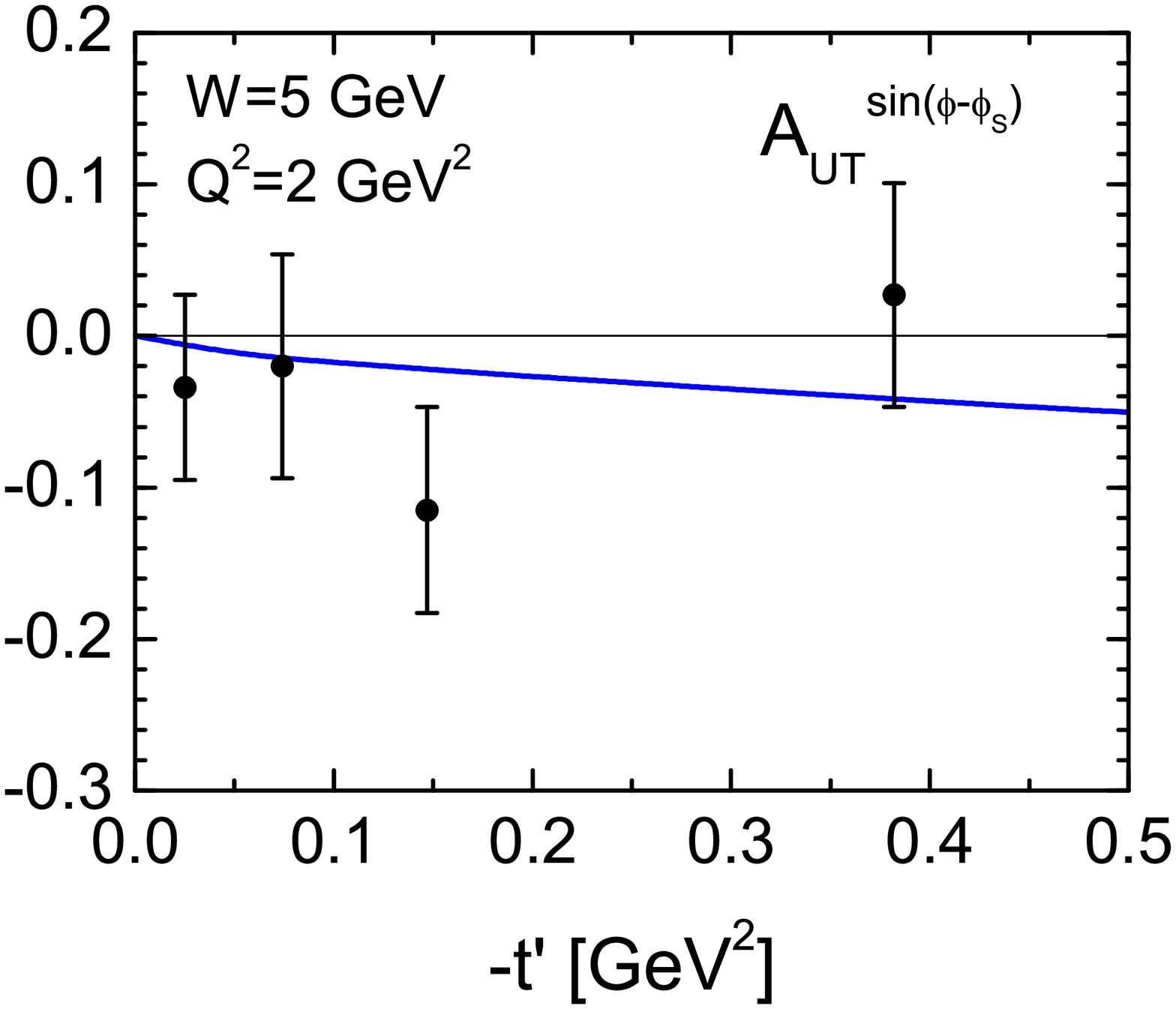}&
\includegraphics[width=6.1cm,height=5cm]{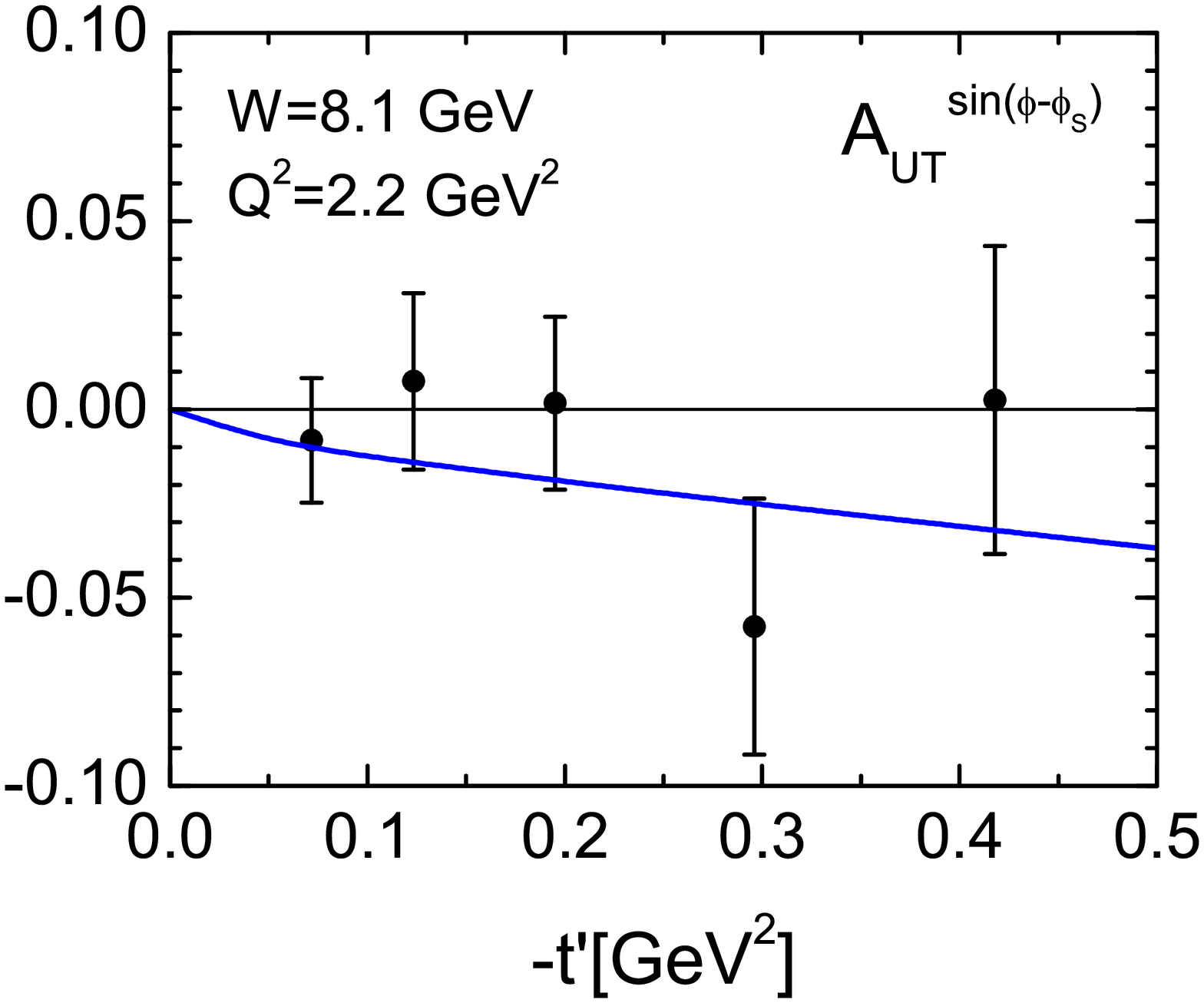}
\end{tabular}
\end{center}
\caption{Model results for the $A_{UT}^{\sin(\phi-\phi_s)}$
asymmetry.
 Left: at HERMES. Right: at COMPASS energy. Data are from
\cite{rostom,autcomp13}.}
\end{figure}

 The $\sin (\phi_s)$ moment of the $A_{UT}$ asymmetry is determined by
the $H_T$ GPDs.

\begin{equation}\label{sinfs}
 A_{UT}^{\sin(\phi_s)} \sim \mbox{Im}[M_{0-,++}^*
M_{0+,0+}]; \;\;\;M_{0-,++}=<H_T>
\end{equation}

\begin{figure}[h!]
\begin{center}
\begin{tabular}{cc}
\includegraphics[width=6.1cm,height=5cm]{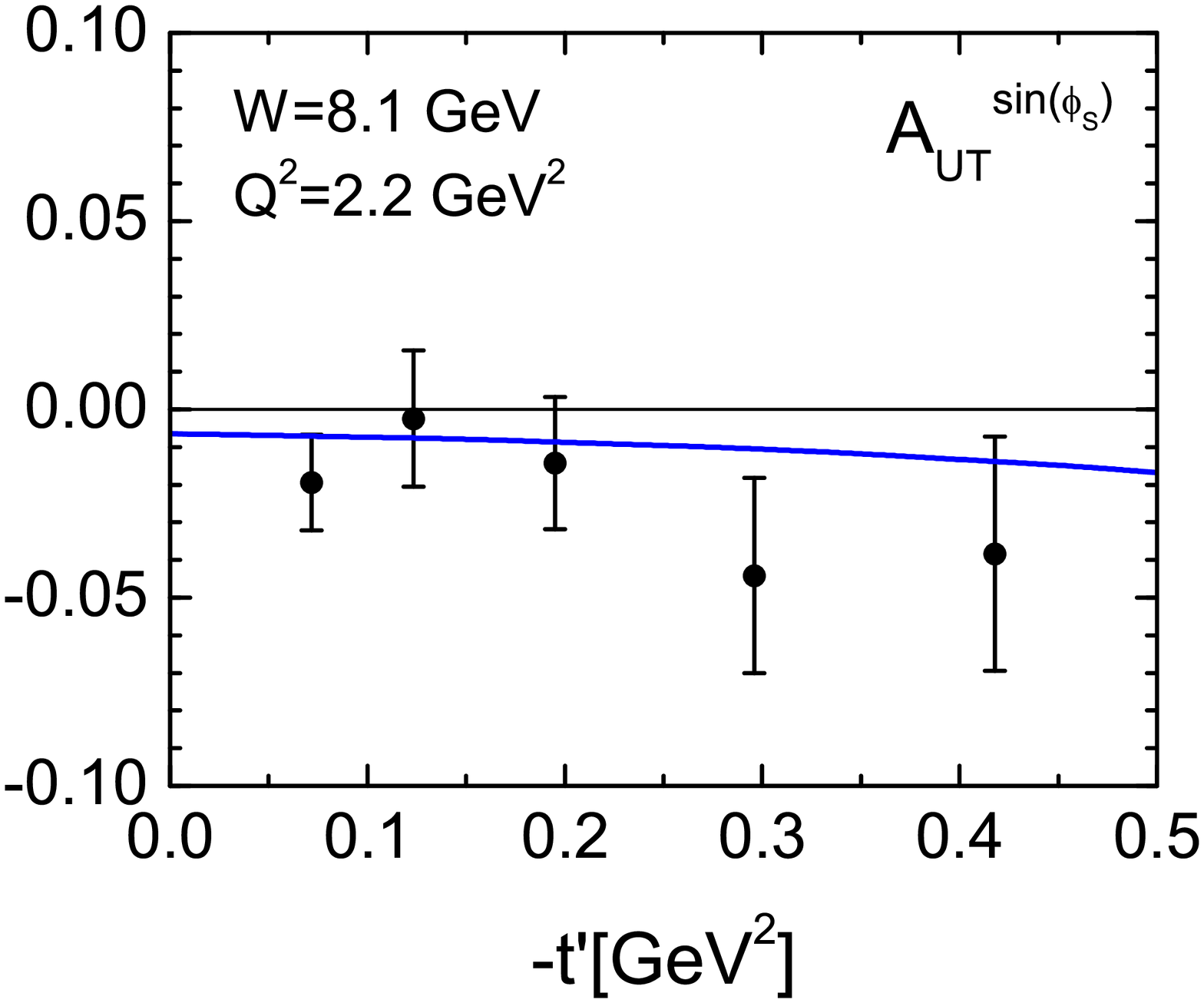}&
\includegraphics[width=6.1cm,height=5cm]{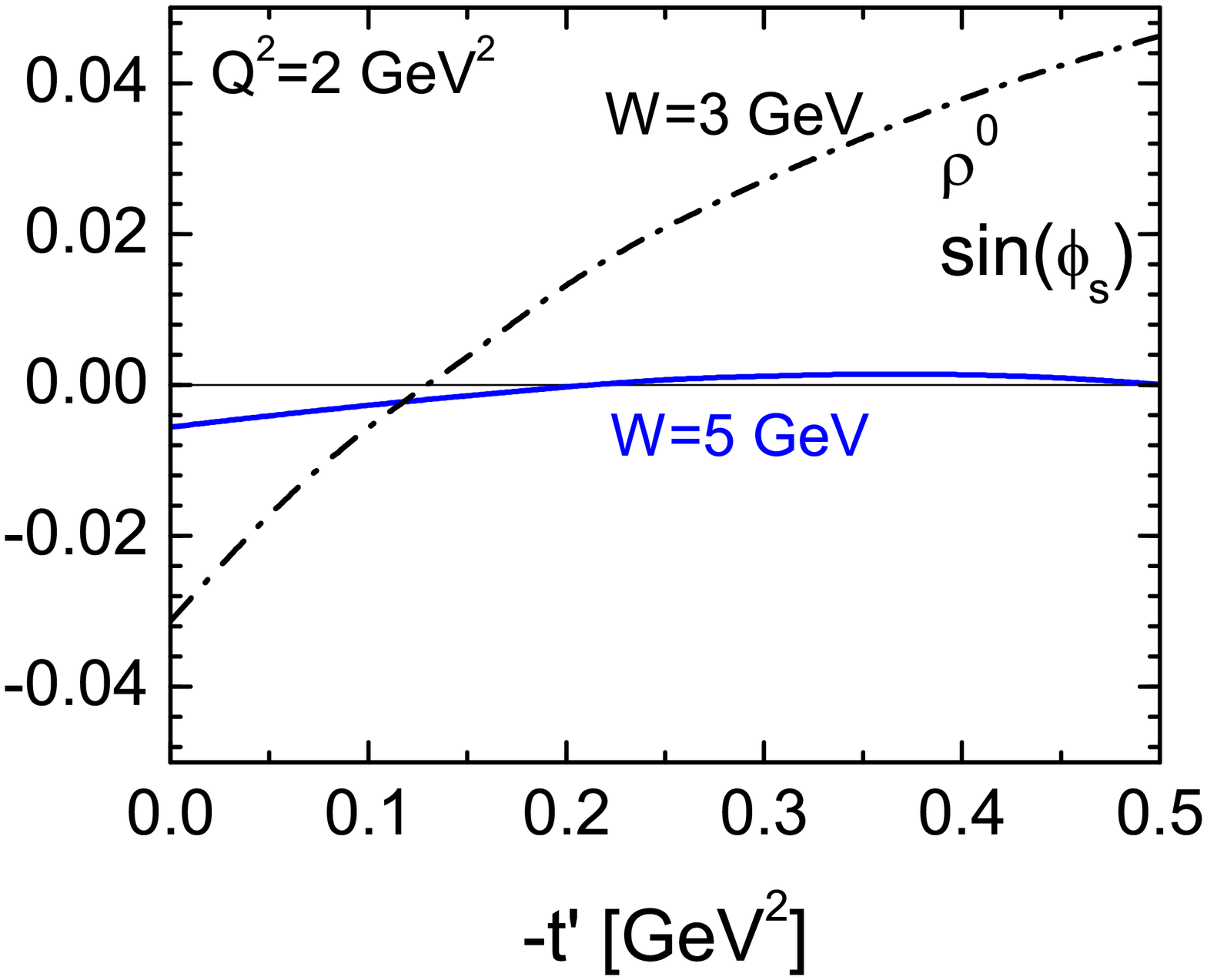}
\end{tabular}
\end{center}
\caption{Left: $A_{UT}^{\sin (\phi_s)}$ asymmetry as COMPASS. Data
are from \cite{autcomp13}. Right: Predicted $A_{UT}^{\sin
(\phi_s)}$ asymmetry at HERMES and CLAS energies.}
\end{figure}

This asymmetry is found to be not small at COMPASS  \cite{gk13}
and compatible with the data \cite{autcomp13} Fig 6 (left). The
energy dependence of $A_{UT}^{\sin(\phi_s)}$ from CLAS to HERMES
is quite rapid and shown in Fig. 6 (right). This prediction can be
verified in a future CLAS experiment to test the $x$- dependence
of GPDs $H_T$.

\begin{figure}[h!]
\begin{center}
\begin{tabular}{cc}
\includegraphics[width=6.1cm,height=5cm]{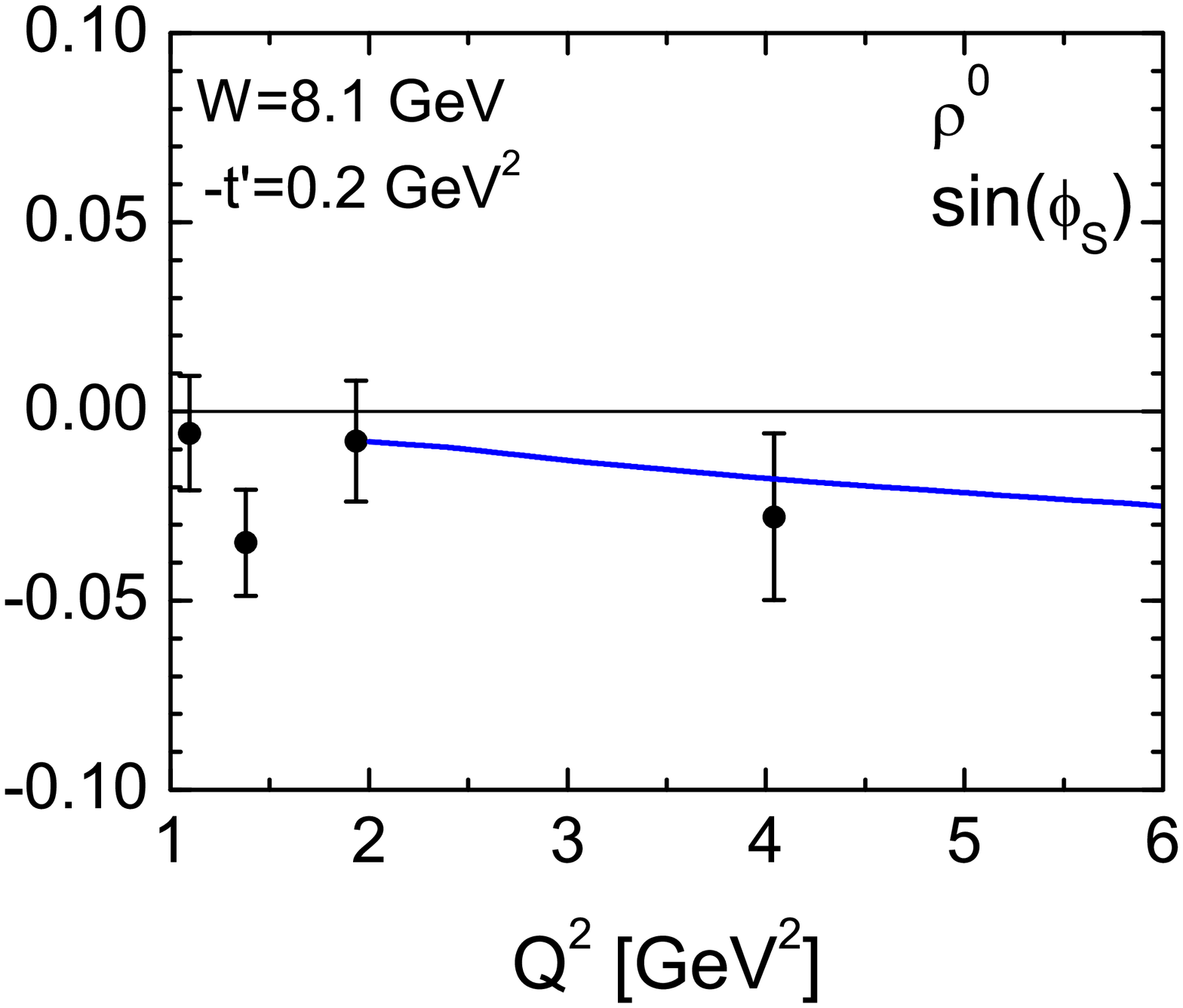}&
\includegraphics[width=6.1cm,height=5cm]{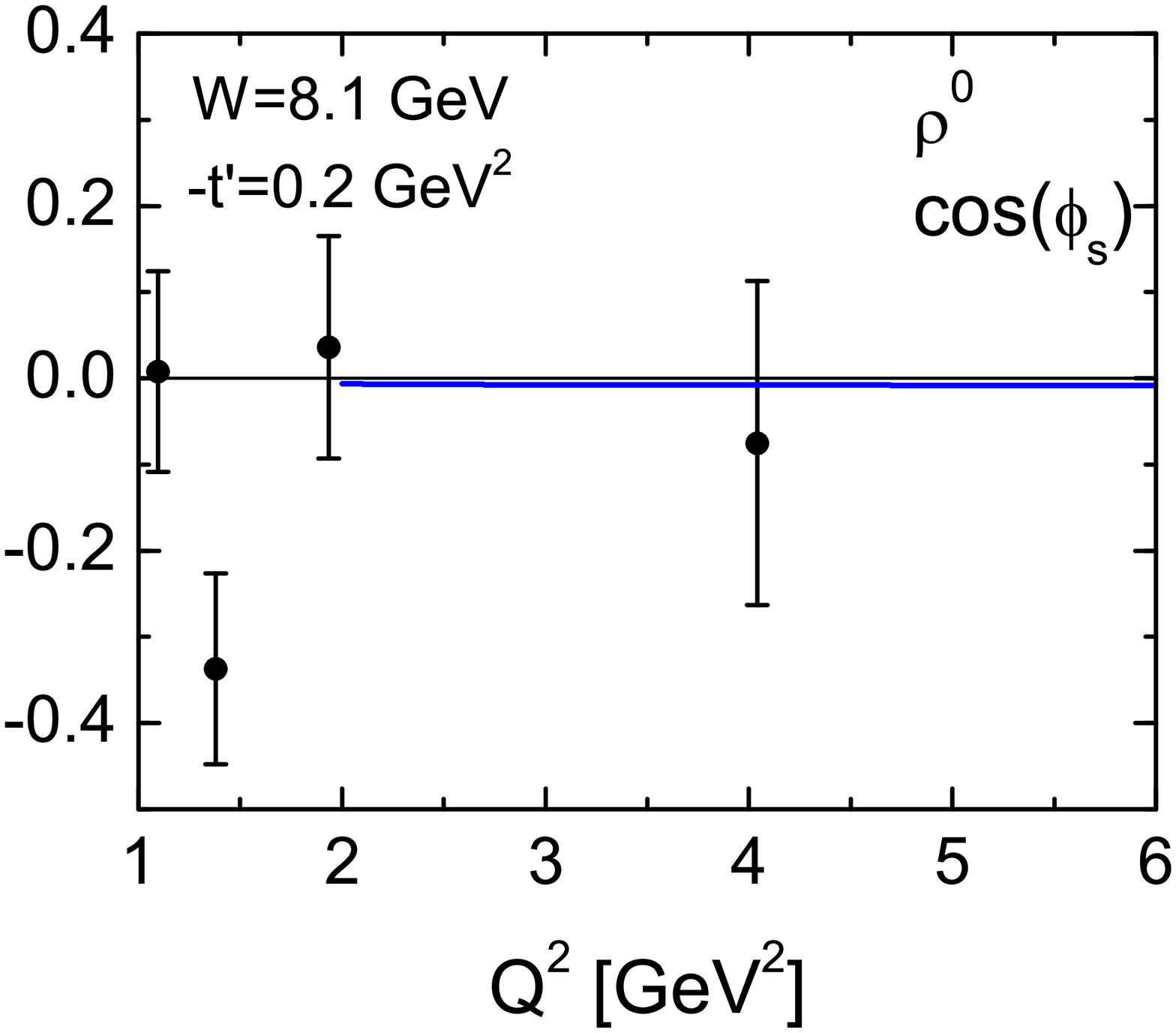}
\end{tabular}
\label{fig:4}
\end{center}
\caption{$Q^2$ dependences of Left: $A_{UT}^{\sin(\phi_s)}$
asymmetry.  Right: $A_{LT}^{\cos(\phi_s)}$ asymmetry at COMPASS
together with data \cite{autcomp13}}
\end{figure}

In Fig.7, we show the $Q^2$ dependencies of
$A_{UT}^{\sin(\phi_s)}$ and $A_{LT}^{\cos(\phi_s)}$ which is
determined by a similar to (\ref{sinfs}) equation only with the
replacement of the imaginary to the real part there. The model
results are close to experimental data.

\section{Conclusion}
The handbag approach, where the amplitudes factorize into the hard
subprocesses and  GPDs \cite{fact}, was successfully applied to
light meson production. The results based on this approach on
cross sections and various spin observables were found to be in
good agreement with data at HERMES, COMPASS and HERA energies at
high $Q^2$ \cite{gk06}.

At the leading-twist accuracy the PM production  is only sensitive
to the GPDs $\widetilde{H}$ and $\widetilde{E}$ which contribute
to the amplitudes for longitudinally polarized virtual photons. It
was found that the leading twist contributions are not sufficient
to describe spin observables in PM production at sufficiently low
photon virtualities $Q^2$. We observe that the experimental data
on the PM leptonproduction also require  contributions from the
transversity GPDs from $H_T$ and $\bar E_T$. Within the handbag
approach the transversity GPDs are accompanied by  twist-3 meson
distribution amplitudes. These transversity contributions provide
large transverse cross sections for most of the pseudoscalar meson
channels \cite{gk11}. There is some indication that large
transversity effects are available now at CLASS \cite{bedl}. Thus,
the transversity GPDs are extremely essential in understanding
spin effects in the PM production.

The role of transversity GPDs in the VM leptoproduction was
investigated within  the handbag approach \cite{gk13}. The
transversity GPDs in combination with twist-3 meson wave functions
occur in the amplitudes with the transversely polarized virtual
photon and a longitudinal polarized vector meson. The importance
of the transversity GPDs was examined in  the SDMEs and in
asymmetries measured with a transversely polarized target.  The
SDMEs  for the light VM production were found to be in good
agreement with HERMES experimental data on the $\rho^0$ production
\cite{airap}. We also estimated the $A_{UT}^{\sin(\phi-\phi_s)}$
transverse target spin asymmetry \cite{gk13}. The results are
consistent with HERMES and COMPASS data \cite{rostom,autcomp13}.
The $A_{UT}^{\sin (\phi_s)}$ asymmetry is found in the model to be
not small at COMPASS \cite{gk13} and also compatible with the data
\cite{autcomp13}. Our predictions were compared with the COMPASS
experimental data in the COMPASS paper \cite{autcomp13}.

We described well the cross section and spin observables for
various meson productions. Thus, we can conclude that the
information on GPDs discussed above should not be   far from
reality. Future experimental results at COMPASS, JLAB12 can give
important information on the role of transversity effects in these
reactions.
\bigskip

This work is supported  in part by the Russian Foundation for
Basic Research, Grant  12-02-00613  and by the Heisenberg-Landau
program.


\begin{thebibliography}{99}
\bibitem{gk06} S.V. Goloskokov, P. Kroll,
  Euro. Phys. J.  C42, 281-301 (2005); ibid C{\bf 50} 829, (2007);
  ibid C{\bf 53} 367, (2008)
  ; ibid C{\bf 59}, 809-819 (2009).
\bibitem{fact} X. Ji, Phys. Rev. \textbf{D55}, (1997) 7114;\\
A.V. Radyushkin,  Phys. Lett. \textbf{ B380}, (1996)  417;\\
J.C.  Collins, et al., Phys. Rev. \textbf{D56}, (1997) 2982.
\bibitem{sterman} J. Botts and G. Sterman,
 Nucl. Phys. \textbf{B325}, (1989) 62.
\bibitem{gk09}  S.V.Goloskokov, P.Kroll, Euro. Phys. J.  \textbf{C65}, (2010)
137.
\bibitem{gk11}  S.V.Goloskokov, P.Kroll, Euro. Phys. J. \textbf{A47}, (2011) 112.
\bibitem{gk13} S.V Goloskokov, P. Kroll,  arXiv: 1310.1472 [hep-ph] (2013)
\bibitem{koerner} R.\ Jakob,  P.\ Kroll,
Phys.  Lett. \textbf{B315},  (1993) 463;\\
J.\ Bolz, J.G.\ K\"orner, P.\ Kroll, Z.\ Phys. \textbf{  A350},
(1994) 145.
  \bibitem{CTEQ6} J. Pumplin, et al.,
 JHEP \textbf{0207}, (2002) 012.
\bibitem{pauli} M. Diehl, T. Feldmann, R. Jakob and P.~Kroll,
  Euro. Phys. J.   \textbf{C39}, (2005) 1.
\bibitem{mus99} I.V. Musatov and A.V. Radyushkin,
  Phys. Rev. \textbf{D61}, (2000) 074027.
\bibitem{ans} M. Anselmino, M.
Boglione, U. D'Alesio, A. Kotzinian, F. Murgia, A. Prokudin and S.
Melis,
  Nucl.  Phys.  Proc.  Suppl.   \textbf{191}, (2009) 98.
\bibitem{lat} M. Gockeler {\it et al.}
[QCDSF Collaboration and UKQCD Collaboration],
  Phys. Rev. Lett.  \textbf{98}, (2007) 222001.
\bibitem{frankfurt99} L.~L.~Frankfurt, P.~V.~Pobylitsa, M.~V.~Polyakov and M.~Strikman,
  Phys.\ Rev.\  \textbf{D60}, (1999) 014010.
\bibitem{bedl} I. Bedlinskiy, et al. (CLAS Collaboration)
 Phys. Rev. Lett. \textbf{109}, (2012) 112001.
\bibitem{vkubar} V.Kubarovsky et al., Proc. of DSPIN-11, Dubna,
September 2011, p. 258.
\bibitem{sgspin11} S.V. Goloskokov. Proc. of DSPIN-11, Dubna,
September 2011, p. 71.
\bibitem{airap} A. Airapetian et al.
(HERMES Collab.),  Euro. Phys. J.   \textbf{C62}, (2009) 659.
\bibitem{rostom} A. Rostomyan et al.
(HERMES Collab.),  arXiv: 0707.2486 [hep-ex].
\bibitem{autcomp13} C. Adolph, et al. (COMPASS Collaboration)
 arXiv: 1310.1454 [hep-ex]  (2013).

\end{thebibliography}
\end{document}